# Analyzing the inter-domain vs intra-domain knowledge flows

**Authors:** Giovanni Abramo[1], Ciriaco Andrea D'Angelo[2]

**Affiliations:**

[1] Universitas Mercatorum, Laboratory for Studies in Research Evaluation, Rome, Italy
ORCID: 0000-0003-0731-3635 - giovanni.abramo@unimercatorum.it

[2] University of Rome "Tor Vergata", Dept of Engineering and Management, Rome, Italy
ORCID: 0000-0002-6977-6611 - dangelo@dii.uniroma2.it

## Abstract
Similar to how innovations often find success in fields other than their original domains, in this study we explore whether the same holds true for scientific discoveries. We investigate the flow of knowledge across scientific disciplines, focusing on connections between citing and cited publications. Specifically, we analyze the connections among cited publications from 2015 indexed in the Web of Science and their citing counterparts to measure rates of knowledge dissemination within and across different fields. Our study aims to address key research questions concerning the disparities between inter- and intra-domain knowledge dissemination rates, the correlation between knowledge dissemination types and scholarly impact, as well as the evolution of knowledge dissemination patterns over time. These findings deepen our understanding of knowledge flows and offer practical insights with significant implications for evaluative bibliometrics.

## 1. Introduction

In the realm of technology innovation, there are numerous examples of inventions that have found success in domains other than the ones for which they were originally created. This phenomenon is often referred to as technology transfer or cross-domain innovation. Here are a few examples.

The adhesive used in Post-it Notes, invented by 3M engineer Spencer Silver, was initially considered a failure because it was not strong enough. However, it turned out to be perfect for creating removable notes, and Post-it Notes became a widely used office product. The microwave oven was originally developed as part of radar technology during World War II. It was not until after the war that it found its way into kitchens for cooking food quickly. Originally developed to treat muscle spasms, Botox (botulinum toxin) is now widely used for cosmetic purposes to reduce wrinkles. Originally developed to treat angina (chest pain associated with heart disease), Viagra (sildenafil) was found to have an unexpected side effect—enhancing male erections. It has since become one of the most well-known medications for treating erectile dysfunction.

These examples illustrate the potential for inventions to find applications across diverse domains, often resulting in unexpected and successful outcomes. Similarly, transitioning from technological research results to scientific ones, numerous cases of cross-fertilization and the emergence of new domains can be identified. Below are some examples.

Between computer science and bioinformatics, the utilization of machine learning and artificial intelligence in genomics and proteomics research (Mann, Kumar, Zeng, & Strauss, 2021). Between physics and biology, the application of statistical physics in modelling biological systems and understanding complex biological processes (Sella & Hirsh, 2005). Between computer science and neuroscience, the use of algorithms inspired by neural networks to simulate and understand brain function (Fong, Scheirer, & Cox, 2018). Between biology and economics, the application of ecological and evolutionary principles in the study of economic systems, such as game theory and evolutionary economics (Tilman, Plotkin, & Akçay, 2020). Between mathematics and medicine, the application of mathematical modeling in epidemiology to predict and control the spread of diseases (Kretzschmar & Wallinga, 2010).

These examples demonstrate how collaboration and knowledge exchange between different scientific domains can lead to breakthroughs and advancements that may not have been possible within the confines of a single discipline. These, along with other specific case studies available in the literature (Douard, Samet, Giakos, & Cavallucci, 2023; Douard, Samet, Giakos, & Cavallucci, 2022; Li, Wang, Yu, & Heng, 2021; Halse & Bjarnar, 2014), represent the tip of an iceberg for the rest that remains unknown. Few studies have investigated the knowledge flows across the domains within the field of physics (Sinatra et al., 2015; Sun & Latora, 2020), or within and across few fields (Yan, 2016). To the best of our knowledge, there are no comprehensive and general studies on the phenomenon globally. In other words, if one were to inquire whether and to what extent each scientific publication has a scholarly impact that transcends the boundaries of its own domain, it is not possible to find an answer in the literature. The question may be relevant for policymakers interested in understanding, for example, whether certain initiatives aimed at fostering a more interconnected and dynamic research landscape are effective. In line with a growing trend to encourage researchers to break down silos, collaborate across disciplines, and leverage the strengths of different fields to address complex challenges and discover innovative solutions.

Utilizing bibliometric techniques, this study aims to contribute to filling this gap. To achieve this objective, we will differentiate between inter-domain knowledge flows and intra-domain knowledge flows. The former pertains to the transfer, exchange, or movement of knowledge across different domains, disciplines, or fields. This implies disseminating knowledge across boundaries and its application or integration into areas beyond its original domain. Conversely, the latter refers to knowledge flows within the same domain, discipline, or field, involving the sharing and transfer of knowledge researchers operating within the same areas of expertise.

Our study is grounded in a fundamental postulate of citation-based bibliometrics (normative theory of citing), wherein a citation represents, net of exceptions (Glänzel, 2008), recognition of the influence of the cited publication on the citing one (Mulkay, 1976; Bloor, 1976; Merton, 1973).

Through the examination of cited-citing relationships within scientific literature and the categorization of their corresponding disciplinary fields, it becomes feasible to quantify the rates of inter- and intra-domain knowledge dissemination associated with each scientific publication, in aggregate, by field, and longitudinally. Specifically, our aim is to address three research questions:

RQ1: What disparities exist between rates of inter- and intra-domain knowledge dissemination, and how do these variances fluctuate across different fields?

RQ2: What correlation exists between the type of knowledge dissemination stemming from a publication and its overall scholarly impact?

RQ3: How do the knowledge dissemination patterns evolve over time?

In the next section, we will elucidate the methodologies employed and the analysis dataset. In Section 3, we will present the results of our analyses. The concluding section will provide our final considerations and potential future investigations.

## 2. Data and methods

Our dataset comprises 2015 global publications indexed in the WoS Core Collection by Clarivate™. For each publication, we examine citing references as of December 31st of each year up to 2022, resulting in eight datasets with citation windows ranging from 0 to 7 years. Regarding field classification, we utilize the subject categories (SCs) of the WoS classification schema, which consists of 254 SCs grouped into 13 scientific macro-areas. Clarivate employs this schema for classifying journals and publication venues. We assign each publication, whether cited or citing, to the SC of its hosting source. In the case of publications hosted in multi-category sources, we assign them to one SC, among those of the hosting source, based on the following criteria:

- The predominant SC among all citing publications (up to December 31st, 2022) for cited publications.
- The predominant SC among publications cited in the bibliography for citing publications.

In the event of a tie, the assignment is randomly chosen among the tied SCs of the hosting source.

In order to assess the magnitude of extra-domain flows, we will measure both the incidence of citing publications assigned to SCs different from that assigned to the cited publication, and the number of distinct SCs assigned to citing publications.

As an illustrative example, consider the publication dx.doi.org/10.1016/j.joi.2015.07.003. According to Clarivate$^{TM}$ data, it received 51 citations as of December 31, 2022. Among the two SCs associated with the journal ("Journal of Informetrics"), analysis of the SCs attributable to the 51 citing publications indicates "Information Science & Library Science" as the most recurrent (24 cases) compared to "Computer Science, Interdisciplinary Applications" (only 4 cases). Therefore, the new knowledge inherent in this publication is assigned to the SC "Information Science & Library Science" and generates intra-domain flows for 47.1% of the total (24 out of 51 citing publications are assigned to this SC). The complementary 52.9% indicates the magnitude of extra-domain flows generated. These flows, in particular, involve 20 distinct SCs, 17 of which are characterized by only one citation. The remaining 10 extra-domain citations concern "Computer Science, Interdisciplinary Applications" (4 citations); Education & Educational Research (2 citations); "Multidisciplinary Sciences" (4 citations).

For analytical purposes, we will also correlate the magnitude of intra- and extra-domain flows with the scholarly impact of each 2015 cited publication. Consistent with Abramo, Cicero, and D'Angelo (2012),[1] scholarly impact will be measured by rescaling the received citations to the average citations of all 2015 cited publications within the same SC.

## 3. Results

The publications indexed in WoS in 2015 and with at least one citation received by 31/12/2022 amount to 2,045,768, distributed by area as shown in Table 1. Columns 2-4 of the table display the distribution of publications according to the type of citation flows generated. Specifically, 9% of the total publications receive citations solely from publications within their own SC (column 3 of the last row). This share varies by area, ranging from a minimum of 1.8% for publications in the area of "Multidisciplinary Sciences" followed by Chemistry (3.5%), to a maximum of 28.2% in "Art and Humanities", followed by "Mathematics" (26.9%).

Conversely, 17.2% of the total publications generate citation flows completely outside their domain, being cited by publications from SCs different from that of the cited publication. This data also sees the primacy of "Art and Humanities" (29.5%), but now followed by "Multidisciplinary Sciences" (29.3%).

"Art and Humanities" indeed presents several distinctive features: the fifth column of Table 1 indicates that in this area, 43.9% of the publications are characterized by a proportion of intra-domain citations greater than 50%, a value second only to that recorded in "Mathematics" (53.6%). Obviously, the data in column 5 (e.g., the 43.9% of "Art and Humanities") also includes that of column 3 (and thus the 28.2% of "Art and Humanities") of cited totally intra-domain.

Overall, the proportion of publications generating predominantly intra-domain flows is 29.1%; below this value, we find four areas: Psychology (16.4%), Biomedical Research (21.7%), Chemistry (22.7%), Biology (23.5%), and, of course, Multidisciplinary Sciences (3.6%). The last column of Table 1 refers to the average value of the share of intra-domain citations over the total; overall, this share averages 38.3%. Therefore, on the totality of citations received by publications in the dataset, slightly less than two-thirds come from citing publications in SCs different from that attributed to the cited publication.

***Table 1: The breakdown of 2015 WoS publications by the type of citations flows,\* by area***

| Area | Total publications | Of which cited totally intra-domain | Of which cited totally extra-domain | Of which predominantly cited intra-domain | Avg share of intra-domain citations over the total |
|---|---|---|---|---|---|
| Art and Humanities | 50310 | 28.2% | 29.5% | 43.9% | 48.9% |
| Biology | 233571 | 5.3% | 14.7% | 23.5% | 34.3% |
| Biomedical Research | 184564 | 5.6% | 16.9% | 21.7% | 33.6% |
| Chemistry | 149608 | 3.5% | 9.8% | 22.7% | 35.6% |
| Clinical Medicine | 421889 | 7.8% | 17.2% | 28.7% | 37.6% |
| Earth and Space Sciences | 98011 | 4.8% | 9.7% | 34.1% | 42.0% |
| Economics | 59921 | 11.1% | 17.8% | 37.5% | 43.9% |
| Engineering | 422690 | 10.6% | 18.8% | 30.9% | 40.3% |
| Law, political and social sciences | 104519 | 14.9% | 24.1% | 34.2% | 40.8% |
| Mathematics | 54589 | 26.9% | 14.8% | 53.6% | 56.8% |
| Multidisciplinary Sciences | 66911 | 1.8% | 29.3% | 3.6% | 12.2% |
| Physics | 176601 | 11.5% | 15.6% | 37.2% | 44.4% |
| Psychology | 22584 | 4.2% | 19.2% | 16.4% | 29.0% |
| Total | 2045768 | 9.0% | 17.2% | 29.1% | 38.3% |

*\* intra-domain refers to citations coming from publications in the same SC as the cited one; extra-domain refers to citations coming from publications in an SC different from that of the cited one*

Descending to the individual SC level, Table 2 presents the top 10 and bottom 10 SCs in terms of intra-domain flow shares. Specifically, the fourth column indicates the average value of the share of citations coming from the same SC as the cited publications, while the fifth column represents its complement to one, namely the average of extra-domain flows. At the top of the table, we find "Literary Reviews". Out of the 276 publications in this SC, 248 generate only extra-domain flows, being cited exclusively by publications from other SCs, hence the highest average value of extra-domain citations at 91.4%. On the opposite end, we have Astronomy & Astrophysics: out of the 16,244 publications in this SC, less than 3% generate totally extra-domain flows, while nearly 39% generate totally intra-domain flows. Overall, the average value of intra-domain citations in Astronomy & Astrophysics is the highest at 82.0%.

*Table 2: Subject categories with the min-max difference in intra vs extra domain flows*

| Subject category | Area | Obs | Intra-domain | Extra-domain | Diff. |
|---|---|---|---|---|---|
| Literary Reviews | Art and Humanities | 276 | 8.6% | 91.4% | -82.8% |
| Psychology, Biological | Psychology | 367 | 9.0% | 91.0% | -82.1% |
| Multidisciplinary Sciences | Multidisciplinary Sciences | 61112 | 9.8% | 90.2% | -80.4% |
| Biophysics | Biology | 2835 | 9.9% | 90.1% | -80.2% |
| Biology | Biology | 7137 | 11.7% | 88.3% | -76.5% |
| Neuroimaging | Clinical Medicine | 557 | 11.9% | 88.1% | -76.2% |
| Medicine, Research & Experimental | Biomedical Research | 18761 | 12.1% | 87.9% | -75.7% |
| Limnology | Earth and Space Sciences | 385 | 12.9% | 87.1% | -74.2% |
| Psychology | Psychology | 1342 | 13.5% | 86.5% | -73.0% |
| Andrology | Clinical Medicine | 433 | 15.2% | 84.8% | -69.6% |
| … | | | | | |
| Education & Educational Research | Law, political and social sciences | 24413 | 59.8% | 40.2% | 19.5% |
| Music | Art and Humanities | 1376 | 60.4% | 39.6% | 20.8% |
| Ophthalmology | Clinical Medicine | 11283 | 60.9% | 39.1% | 21.9% |
| Philosophy | Art and Humanities | 5678 | 61.1% | 38.9% | 22.2% |
| Classics | Art and Humanities | 1033 | 61.4% | 38.6% | 22.9% |
| Religion | Art and Humanities | 5842 | 63.4% | 36.6% | 26.8% |
| Law | Law, political and social sciences | 8379 | 65.7% | 34.3% | 31.5% |
| Physics, Particles & Fields | Physics | 9339 | 66.0% | 34.0% | 32.0% |
| Mathematics | Mathematics | 23056 | 75.7% | 24.3% | 51.4% |
| Astronomy & Astrophysics | Physics | 16244 | 82.0% | 18.0% | 63.9% |

From these initial analyses, there appears to be greater extra-domain flows than intra-domain ones. Table 3 somewhat confirms this: only in 31 out of the total 254 SCs (12.2% of cases) is there an average share of intra-domain citations greater than 50%. At the subject area level, this occurs in 9 out of the 27 SCs in Art and Humanities and in 2 out of the 6 SCs in Mathematics, while it does not occur in any of the 14 SCs in Biomedical Research or the 8 SCs in Chemistry.

*Table 3: Number of SC by area where the average share of intra-domain citations è above 50%*

| Area | No. of SCs | Of which the average share of intra-domain citations is above 50% |
|---|---|---|
| Art and Humanities | 27 | 9 (33.3%) |
| Biology | 30 | 3 (10.0%) |
| Biomedical Research | 14 | 0 (0.0%) |
| Chemistry | 8 | 0 (0.0%) |
| Clinical Medicine | 43 | 4 (9.3%) |
| Earth and Space Sciences | 13 | 1 (7.7%) |
| Economics | 12 | 2 (16.7%) |
| Engineering | 42 | 2 (4.8%) |
| Law, political and social sciences | 27 | 3 (11.1%) |
| Mathematics | 6 | 2 (33.3%) |
| Multidisciplinary Sciences | 3 | 0 (0.0%) |
| Physics | 19 | 4 (21.1%) |
| Psychology | 10 | 1 (10.0%) |
| Total | 254 | 31 (12.2%) |

Finally, Table 4 presents the results of the analysis concerning the spread of extra-domain flows generated by the 2015 publications in the dataset. Specifically, for each subject area, descriptive statistics are provided regarding the number of distinct SCs attributed to the citing publications in the dataset, per cited subject area. The distributions are all right-skewed with maximum values practically never lower than 90, a general mean of 6.26, and a median of 5.00. It is evident that this spread tends to differentiate between the areas: as expected, "Multidisciplinary Sciences" has the highest values, with a publication attracting citations from publications in as many as 222 different SCs. This is exemplified by a review published in Nature (10.1038/nature14539) titled "Deep learning," cited by a remarkable 19,428 publications, spreading over 222 different SCs. On the other hand, "Art and Humanities" and "Mathematics" confirm themselves as areas with a lower spread of extra-domain flows.

*Table 4: Spread of the flows generated by cited publications. Descriptive statistics of the number of subject categories from which citations to publications in each area originate*

| Area | Avg | St. dev. | Min | Max | Median | Q1 | Q3 |
|---|---|---|---|---|---|---|---|
| Art and Humanities | 2.48 | 2.58 | 1 | 89 | 2 | 1 | 3 |
| Biology | 7.55 | 6.67 | 1 | 144 | 6 | 3 | 10 |
| Biomedical Research | 7.53 | 6.56 | 1 | 168 | 6 | 3 | 10 |
| Chemistry | 6.86 | 5.17 | 1 | 91 | 6 | 3 | 9 |
| Clinical Medicine | 6.74 | 6.62 | 1 | 219 | 5 | 2 | 9 |
| Earth and Space Sciences | 6.70 | 5.59 | 1 | 171 | 5 | 3 | 9 |
| Economics | 5.69 | 5.97 | 1 | 148 | 4 | 2 | 7 |
| Engineering | 5.36 | 5.35 | 1 | 206 | 4 | 2 | 7 |
| Law, political and social sciences | 4.94 | 5.75 | 1 | 140 | 3 | 1 | 6 |
| Mathematics | 3.51 | 4.67 | 1 | 216 | 2 | 1 | 4 |
| Multidisciplinary Sciences | 9.68 | 8.47 | 1 | 222 | 8 | 4 | 13 |
| Physics | 4.85 | 4.24 | 1 | 96 | 4 | 2 | 6 |
| Psychology | 7.84 | 6.84 | 1 | 90 | 6 | 3 | 10 |
| Total | 6.26 | 6.09 | 1 | 222 | 5 | 2 | 8 |

## 5.1 Association between citation flows generated and overall scholarly impact of a publication.

We now inquire whether there exists a significant association between the type of citation flow generated by a publication (intra- vs extra-domain) and its overall scholarly impact. Figure 1 illustrates the dispersion of values for all publications (4034 in total) in the "Information science and library science" SC. Particularly notable is that as the share of intra-domain flows increases, despite evident fluctuations, the impact generally tends to decrease. The same trend is observed for the 22584 publications in the "Psychology" area (Figure 2).

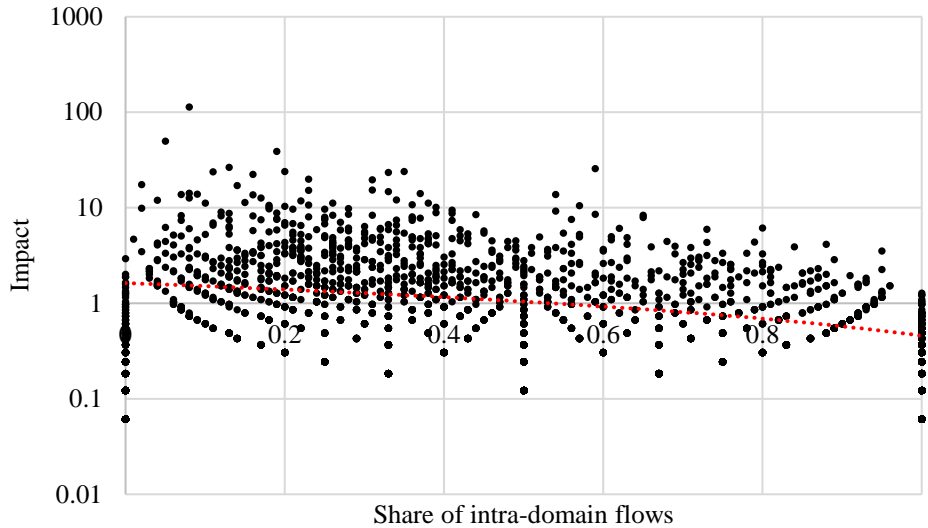

*Figure 1: Scatterplot of impact vs share of intra-domain knowledge flows for the 4034 publications in the subject category "Information science and library science"*

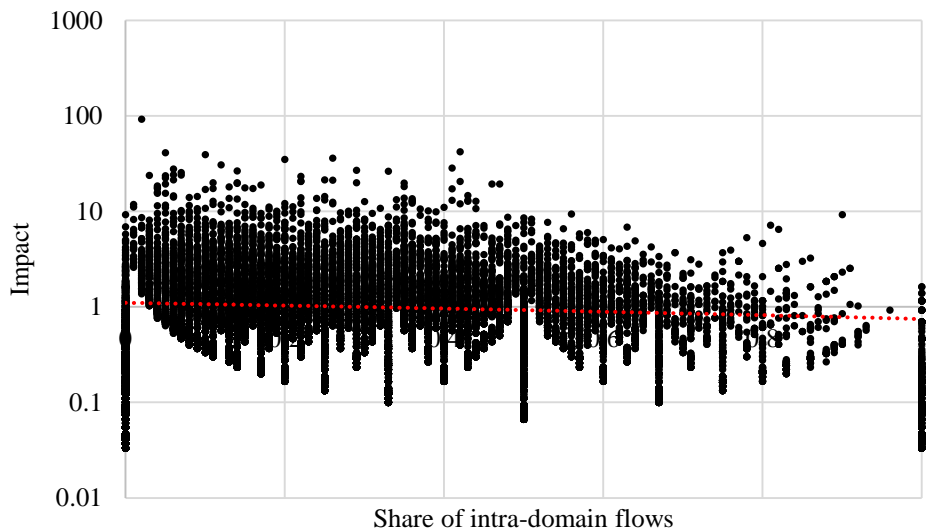

*Figure 2: Scatterplot of impact vs share of intra-domain flows, for 22,584 publications in the area "Psychology"*

Aggregating the data by subject area, Table 5 presents the average impact value for two subsets of publications: those predominantly generating intra-domain flows (share of intra-domain flows > 50%) and those, conversely, predominantly generating extra-domain flows (share of intra-domain flows < 50%). The data clearly indicate that the former have a lower impact compared to the latter, both overall and at the subject area level. Specifically, publications predominantly generating intra-domain flows consistently exhibit an impact below the expected value (which, given how the indicator is constructed, is represented by one). Conversely, those predominantly generating extra-domain flows consistently show an impact exceeding the expected value. The difference between the two means, although consistently significant, is minimal in the "Economics" (0.987 vs 1.091) and "Law, political and social sciences" (0.955 vs 1.094) SCs, and maximal in Multidisciplinary Sciences (0.568 vs 1.031) and Psychology (0.690 vs 1.106).

The presence of a negative association between the incidence of intra-domain flows and impact is confirmed by correlation analyses conducted at the SC level, the results of which are aggregated at the area level in Table 6. Overall, there is a clear prevalence of negative correlation cases (209 out of the total 254 SCs), but also some instances (the remaining 46, accounting for 17.7% of the total) of positive correlation, with a maximum observed in Economics (4 out of the total 12 SCs).

*Table 5: Average impact by type of predominant flow generated*

| Area | Avg impact of predominantly cited intra-domain | Avg impact of predominantly cited extra-domain | Delta impact |
|---|---|---|---|
| Art and Humanities | 0.902 | 1.144 | -0.243 |
| Biology | 0.849 | 1.096 | -0.247 |
| Biomedical Research | 0.853 | 1.094 | -0.241 |
| Chemistry | 0.819 | 1.106 | -0.286 |
| Clinical Medicine | 0.889 | 1.109 | -0.220 |
| Earth and Space Sciences | 0.944 | 1.092 | -0.148 |
| Economics | 0.987 | 1.091 | -0.104 |
| Engineering | 0.907 | 1.129 | -0.223 |
| Law, political and social sciences | 0.955 | 1.094 | -0.139 |
| Mathematics | 0.937 | 1.191 | -0.254 |
| Multidisciplinary Sciences | 0.568 | 1.031 | -0.463 |
| Physics | 0.935 | 1.120 | -0.185 |
| Psychology | 0.690 | 1.106 | -0.416 |
| Total | 0.899 | 1.107 | -0.208 |

*Table 6: Impact vs share of intra-domain flows correlation at SC level, by area*

| Area | N. of SCs | Of which with a negative correlation | Correlation min | Correlation max | Correlation average |
|---|---|---|---|---|---|
| Art and Humanities | 27 | 25 (92.6%) | -0.156 | 0.031 | -0.057 |
| Biology | 30 | 28 (93.3%) | -0.166 | 0.037 | -0.056 |
| Biomedical Research | 14 | 12 (85.7%) | -0.112 | 0.008 | -0.050 |
| Chemistry | 8 | 6 (75.0%) | -0.107 | 0.051 | -0.041 |
| Clinical Medicine | 43 | 33 (76.7%) | -0.179 | 0.027 | -0.042 |
| Earth and Space Sciences | 13 | 12 (92.3%) | -0.129 | 0.009 | -0.049 |
| Economics | 12 | 8 (66.7%) | -0.099 | 0.062 | -0.023 |
| Engineering | 42 | 32 (76.2%) | -0.178 | 0.060 | -0.039 |
| Law, political and social sciences | 27 | 22 (81.5%) | -0.108 | 0.029 | -0.033 |
| Mathematics | 6 | 6 (100.0%) | -0.063 | -0.012 | -0.030 |
| Multidisciplinary Sciences | 3 | 3 (100.0%) | -0.039 | -0.018 | -0.028 |
| Physics | 19 | 13 (68.4%) | -0.133 | 0.095 | -0.020 |
| Psychology | 10 | 9 (90.0%) | -0.096 | 0.019 | -0.043 |
| *Total* | 254 | 209 (82.3%) | -0.179 | 0.095 | -0.042 |

Figure 3 shows how the average impact of publications varies depending on the number of distinct SCs from which the citations they receive originate. There is a clear, positive association between these two variables. When citations come from publications spanning over 7 SCs, the average impact exceeds the expected value (1.098 when the number of impacted subject categories is 8); when they come from publications spanning over 20 SCs, the average impact is at least 5 times greater than the expected value, and it continues to increase nonlinearly. Finally, Figure 4 presents the aggregation results at the SC level and displays their scatterplot, with the average number of impacted SCs (by the cited publications of the SC) on the x-axis and the impact difference between those predominantly cited extra-domain versus intra-domain on the y-axis. There is a clear predominance of cases with positive values on the y-axis (indicating SCs where the average impact of publications predominantly cited extra-domain is greater than that of intra-domain). It is also noticeable that this difference tends to increase with the number of impacted SCs, i.e., distinct SCs attributed to the citing publications.

*Figure 3: Number of cited publications and their average impact by the number of impacted SCs*

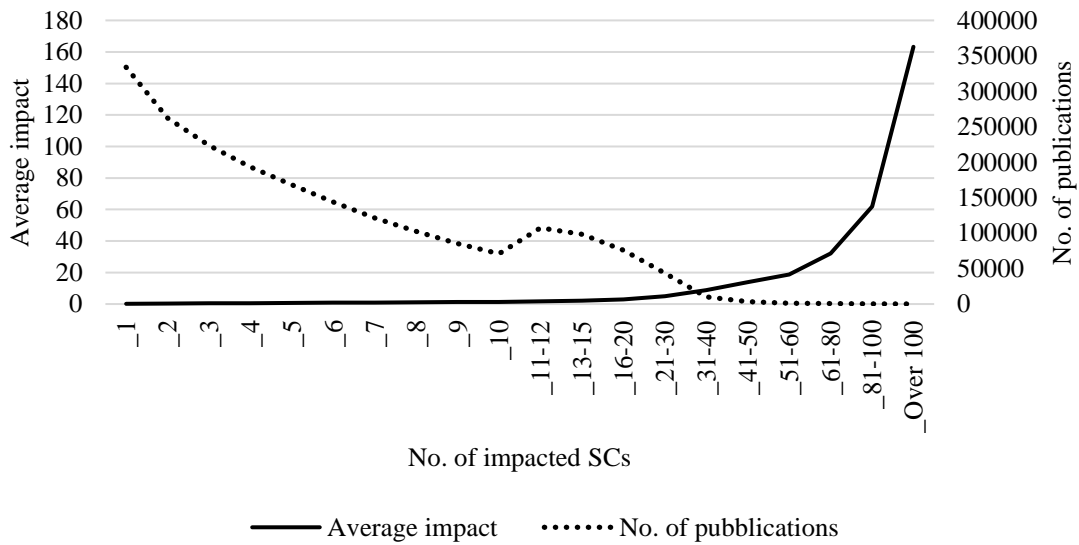

*Figure 4: Scatterplot of SCs by the difference in impact between publications predominantly cited extra-domain versus intra-domain, as a function of the number of SCs affected by citation flows.*

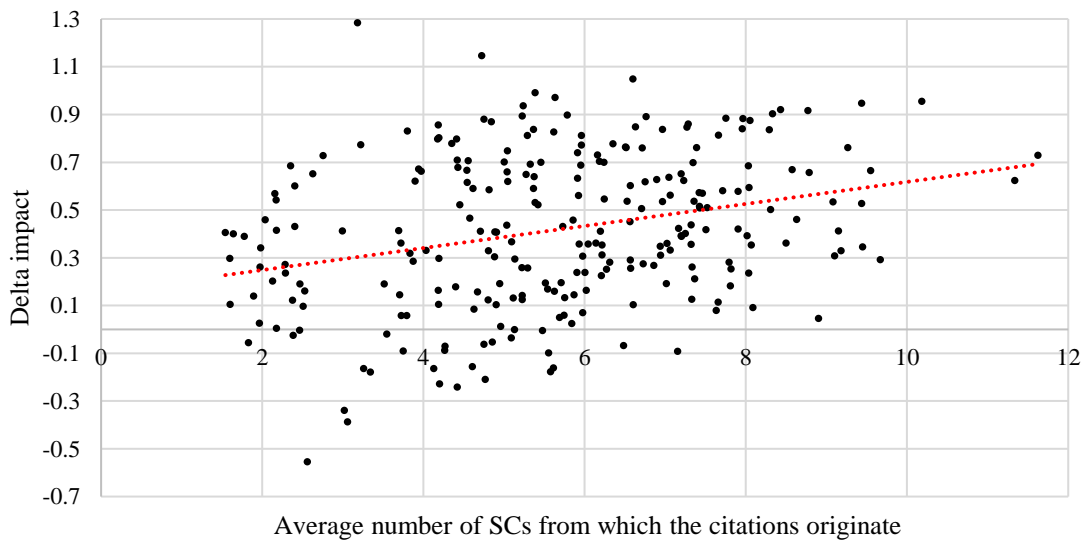

## 5.2 The temporal dynamics of citation flows

We now wonder if the magnitude of intra- and extra-domain flows observed previously considering a citation window of 7 years might be significantly different by reducing the length of the citation window. Specifically, we repeat some of the analyses proposed earlier considering a citation window of only two years, i.e., limiting the dataset to only citing publications with publication years between 2015 and 2017.

Obviously, the first effect that emerges is that the dataset for analysis significantly reduces in size. Figure 5 shows that at an overall level, 19% of the 2015 publications previously considered (i.e., cited at least once by December 31, 2022) had not been cited by December 31, 2017. The difference between areas can be attributed to what is known as immediacy, i.e., the average speed at which knowledge flows originate depending on the specificities of the fields. In Chemistry, 91.9% of the dataset's publications have already received at least one citation within two years of their publication. This proportion decreases drastically in social areas (72.1% in Economics and 66.9% in Law, political and social sciences) and drops below 50% in Art and Humanities (48.4%).

*Figure 5: Share of 2015 publications cited after two calendar years*

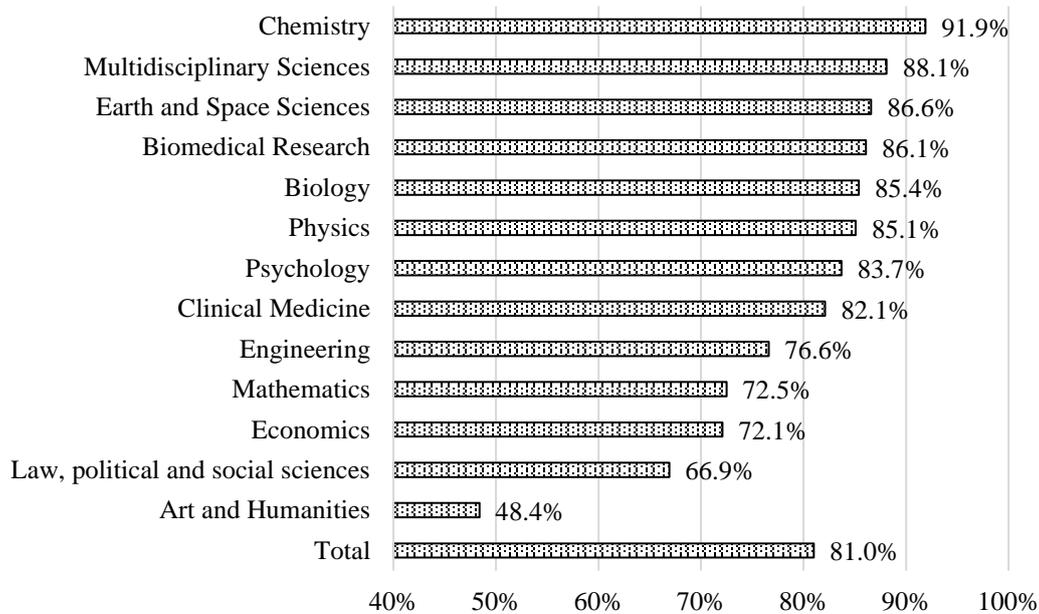

With a citation window of only two years, the proportion of publications cited totally intra-domain is 17.4% overall (Figure 6), nearly double the 9% observed with a 7-year window (third column, last row of Table 1). The difference in share is significant across all areas. In Earth and Space Sciences, it even triples, increasing from 4.8% to 14.8%. As indicated in Figure 7, the proportion of publications predominantly cited intra-domain also increases significantly in this new scenario (33.6% overall) compared to the previous one (29.1%). The difference is evident in all areas except Mathematics.

*Figure 6: Share of 2015 publications cited totally intra-domain after 2 and 7 calendar years*

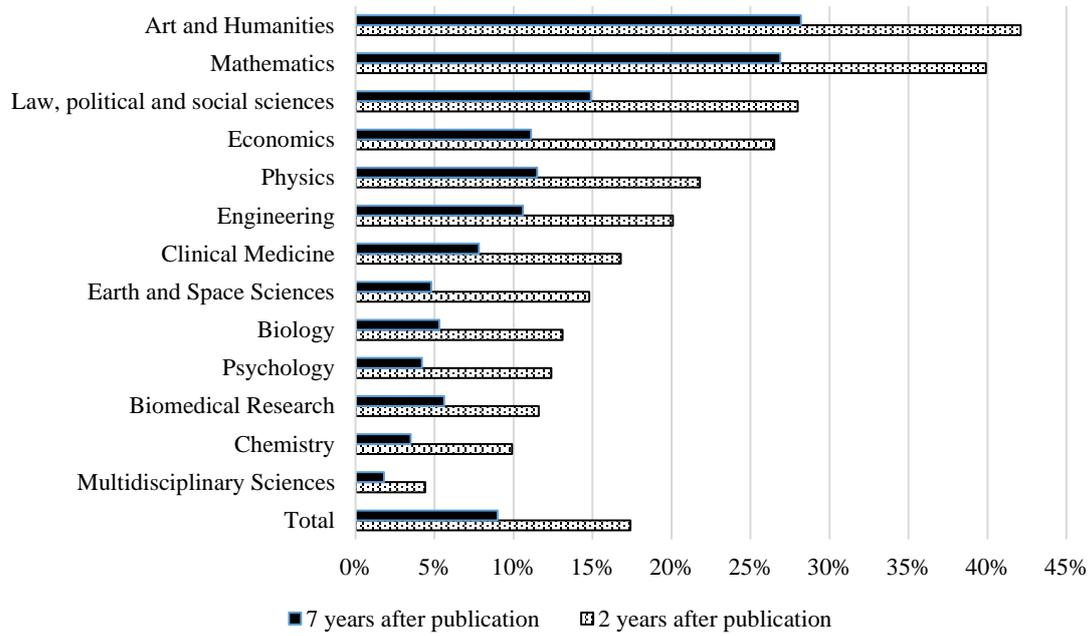

*Figure 7: Share of 2015 publications cited predominantly intra-domain after 2 and 7 calendar years*

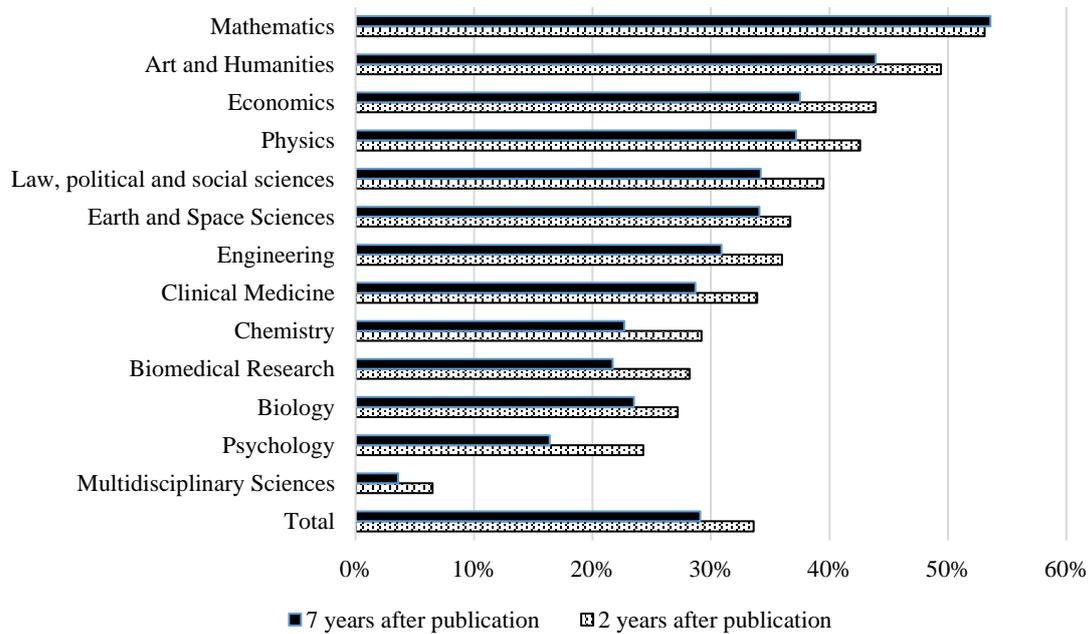

The average value of the share of intra-domain citations increases from the 38.3% observed at 7 years to 41.6% recorded at 2 years, as indicated in Figure 8. In this case, the observed difference remains consistent across all areas, including Mathematics, although in this area, it is very small (56.8% vs. 57.9%).

*Figure 8: Average share of intra-domain citations to 2015 publications, after 2 and 7 calendar years, by area*

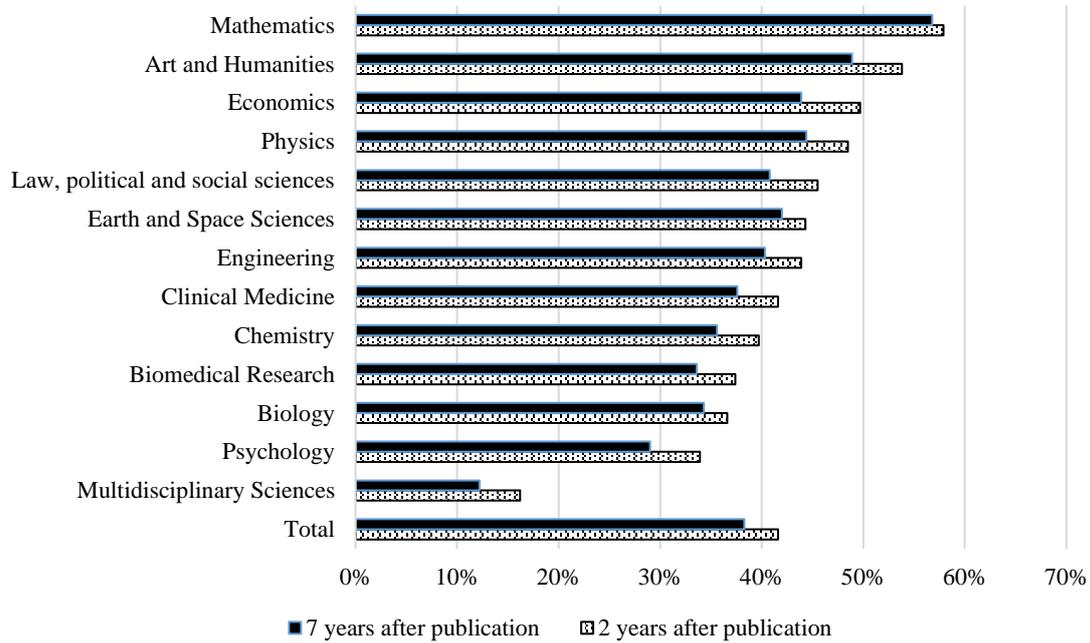

### 4. Conclusions

In this study, adopting a forward-looking perspective and drawing inspiration from the phenomenon of successful innovations in domains different from those for which they were originally created, we investigated the scholarly impact of scientific publications in domains distinct from their own. Specifically, we analyzed over two million publications from 2015 indexed in WoS. As of 31/12/2022, these publications had collectively received nearly 48 million citations from 15.2 million unique citing publications. By assigning each (citing and cited) a specific subject category out of the 254 total categories provided by the WoS classification schema, we were able to analyze the rates of inter- and intra-domain knowledge flows, and how they vary across different areas.

The empirical analysis reveals that 9% of the publications are cited entirely within the same domain, compared to 17.2% that are cited entirely outside their domain. Overall, 29.1% of the 2015 WoS publications are predominantly cited within the same domain, while 38.3% are predominantly cited outside their domain. In other words, scientific publications within a specific SC tend to have a greater impact on the literature of other SCs. This overall trend is consistent across different subject areas, with the exception of "Mathematics", where intra-domain citation flows are more substantial than extra-domain flows. Indeed, only in 12.2% of total SCs, the average share of intra-domain citations is greater than 50%, primarily in Arts and Humanities and Mathematics SCs, while this pattern does not occur in any SCs in Biomedical Research or Chemistry. Arts and Humanities present several distinctive features, perhaps due to their certainly limited coverage in bibliometric databases compared to STEMM areas. Regarding extra-domain citation flows, the analysis shows that overall, a scientific publication generates citation flows to publications belonging to an average of at least six distinct SCs, but with significant variability and anomalous values attributable to publications impacting hundreds of distinct SCs.

The analyses conducted for the second research question reveal a significant association between the type of citation flow generated by a publication and its overall scholarly impact. On average, publications predominantly cited within the same domain exhibit a 20% lower impact compared to those predominantly cited outside their domain. Furthermore, the impact of a publication generally increases significantly with the share of extra-domain citations received and with the number of distinct SCs from which these citations originate.

Moving to the third research question, the data confirm that the minimum time for a publication to receive citations varies across subject areas. Additionally, it is observed that initial citation flows tend to be predominantly within the same domain. Two years from publication, the difference between intra- and extra-domain flows is lower than the observation at seven years. Notably, Mathematics and Arts and Humanities emerge as exceptions in this pattern.

It has been observed that inventions often benefit from integrating knowledge originating from diverse technological domains (Nemet & Johnson, 2012). Numerous instances exist wherein significant inventions result from combinations and transfers across various technological domains (Mowery & Rosenberg, 1998; Ruttan, 2001; Arthur, 2007). The literature on this topic, whose origins can be traced back to Schumpeter (1934), is particularly rich (Hur, 2024; Garcia-Vega, 2006; Verspagen & De Loo, 1999). In the context of scientific literature, it may be equally intriguing to inquire about the extent to which scientific results benefit from discoveries made in diverse scientific domains. This certainly provides an interesting starting point for potential future work that adopts a specular backward perspective.

Moreover, the significant difference between the number of intra-domain and extra-domain flows prompts an important reflection: current national research evaluation exercises (such as the Italian VQR or the UK REF) primarily rely on a peer-review approach. However, given the findings of the proposed study, it begs the question of how peer reviewers, experts in one domain, assess the impact, often unexpected, of a publication in different domains? This question becomes even more urgent at a time when the debate on reforming research evaluation systems seems to be decidedly shifting the focus away from bibliometric approaches. In this regard, the Coalition for Advancing Research Assessment (CoARA)[2] has proposed an agreement, currently with over 600 signatories committed to basing research assessment primarily on qualitative evaluation, with peer review being central. In light of the results obtained in this study, the authors express significant concern and hope that the work somehow prompts a reflection that currently appears to be lacking even within the community of bibliometricians themselves.

With that being said, the authors recognize that when interpreting the findings of the analysis, it's important to consider the typical limitations, cautions, and assumptions of bibliometrics. Specifically, citations may not always validate the true "utility" of knowledge contained in referenced publications, and they may not encompass all potential knowledge flows. Additionally, the results can be influenced by the classification system used for both the cited and citing publications: alterations to this system will inevitably lead to outcomes that could vary significantly.

---